\documentclass{article}

\usepackage{PRIMEarxiv}

\usepackage[utf8]{inputenc} % allow utf-8 input
\usepackage[T1]{fontenc}    % use 8-bit T1 fonts
\usepackage{hyperref}       % hyperlinks
\usepackage{url}            % simple URL typesetting
\usepackage{booktabs}       % professional-quality tables
\usepackage{amsfonts}       % blackboard math symbols
\usepackage{nicefrac}       % compact symbols for 1/2, etc.
\usepackage{microtype}      % microtypography
\usepackage{lipsum}
\usepackage{fancyhdr}       % header
\usepackage{graphicx}       % graphics
\graphicspath{{media/}}     % organize your images and other figures under media/ folder

\usepackage{tabularx}

\usepackage{xurl}

\title{Probing Experts’ Perspectives on AI-Assisted Public Speaking Training
\thanks{\textit{\underline{Citation}}: 
\textbf{Fourati, N., Barkar, A., Dragée, M., Danthon-Lefebvre, L., Chollet, M. (2025). Probing Experts’ Perspectives on AI-Assisted Public Speaking Training. arXiv preprint arXiv:2507.07930}} 
}

\author{
  Nesrine Fourati \\
  Independent Researcher\\
  France \\
  \texttt{nesrine.fourati@gmail.com} \\
  \And
  Alisa Barkar \\
  LTCI, Telecom Paris\\
  France \\
  \texttt{alisa.barkar@telecom-paris.fr} \\
  \And
  Marion Dragée \\
  LS2N, IMT Atlantique \\
  France \\
  \texttt{marion.dragee@imt-atlantique.fr} \\
  \And
  Liv Danthon-Lefebvre \\
  LS2N, IMT Atlantique \\
  France \\
  \texttt{liv.danthon@imt-atlantique.fr} \\
  \And
  Mathieu Chollet \\
  School of Computing Science \\
  University of Glasgow \\
  Scotland, UK \\
  \texttt{mathieu.chollet@glasgow.ac.uk} \\
}

\begin{document}

\maketitle

\begin{abstract}
\textbf{Background:}
Public speaking is a vital professional skill, yet it remains a source of significant anxiety for many individuals. Traditional training relies heavily on expert coaching, but recent advances in AI has led to novel types of commercial automated public speaking feedback tools. However, most research has focused on prototypes rather than commercial applications, and little is known about how public speaking experts perceive these tools.

\textbf{Objectives:}
This study aims to evaluate expert opinions on the efficacy and design of commercial AI-based public speaking training tools and to propose guidelines for their improvement.

\textbf{Methods:}
The research involved 16 semi-structured interviews and 2 focus groups with public speaking experts. Participants discussed their views on current commercial tools, their potential integration into traditional coaching, and suggestions for enhancing these systems.

\textbf{Results and Conclusions:}
Experts acknowledged the value of AI tools in handling repetitive, technical aspects of training, allowing coaches to focus on higher-level skills. However they found key issues in current tools, emphasising the need for personalised, understandable, carefully selected feedback and clear instructional design. Overall, they supported a hybrid model combining traditional coaching with AI-supported exercises.
\end{abstract}

% keywords can be removed

\section{Introduction}\label{intro}

Communication skills like public speaking are essential in both personal and professional contexts, consistently recognised as one of the 21\textsuperscript{st}-century core competencies \cite{Chan2011,Joynes2019}. These skills contribute to effective interpersonal interaction, leadership, collaboration, and persuasion in increasingly hybrid work environments.

Public speaking training relies on face-to-face sessions with experienced coaches offering personalised feedback and nuanced guidance. However, digital transformation and artificial intelligence (AI) integration into education are reshaping communication skills development. This has sparked research exploring AI tools' potential for public speaking skills learning. Research prototypes developed over the past decade attempted to automate feedback based on gestures, speech features, and non-verbal behaviours \cite{Zhao2017,Zeng2019,Wang2020,Zeng2022,NiebuhrWebPitcher2021}. These systems consistently demonstrated benefits including increased confidence \cite{Huang2024} and improved performance \cite{chollet2022training}.

Meanwhile,  commercial training platforms like Poised, Orai, and Yoodli (see Section~\ref{tools_description} and Table~\ref{tab:tool_analysis}) now promise immediate, automated feedback on presentation dimensions, capitalising on AI technologies including natural language processing, computer vision and audio signal processing. However, little is known about experts' opinions of these training tools. Expert insights remain largely absent from literature investigating commercial AI-driven public speaking tools, despite their unique understanding of effective public speaking pedagogy and ability to assess communicative performances and provide personalised, actionable feedback to trainees.

This research addresses this gap by engaging public speaking experts in interviews and focus groups, inviting critical reflection on selected commercial tools. The study reports on expert perceptions of these systems through which we identify design principles to guide the development of public speaking training solutions better aligned with experts' pedagogical insights. Our work contributes to reconciling AI innovations and pedagogical expertise to advance public speaking skills learning.

The paper is organised as follows. Section~\ref{background} presents research on academic prototypes and an overview of commercial systems. Section~\ref{methodology} presents our methodology and two studies. Section~\ref{study1-itw} describes interviews with 16 coaches on current public speaking coaching practices. Section~\ref{study2-wrskp} details a design workshop study in which we presented systems to 7 coaches and investigated opinions through focus groups and creativity sessions. Section~\ref{discussion} discusses findings, comparing study insights to public speaking literature, and Section~\ref{conclusion} concludes with recommendations.

\section{Background}\label{background}

\begin{table*}[t]
\centering
\caption{Overview of academic systems for public speaking training evaluated only with end-users.}
\footnotesize
\begin{tabular}{p{1.5cm} p{2.8cm} p{3.3cm} p{2.8cm} p{4cm}}
\\
\hline
\textbf{Tool} & \textbf{Description} & \textbf{Feedback} & \textbf{Cues} & \textbf{Evaluation} \\
\hline

\\

%\textbf{AwareMe}  \cite{Bubel2016} & Wearable prompting system for vocal behavior awareness & Haptic and visual feedback & Voice pitch, filler words, and words per minute & \textbf{5 users}; Potential of vocal feedback, review of haptic feedback (awareness of time instead of anxiety notifications)\\ \\

\textbf{AutoManner} \cite{Tanveer2016} & Analyses mannerisms via video & Annotated video feedback with advice & Repetitive gestures & \textbf{27 users}. Increased awareness of body language. \\ \\

\textbf{Cicero (2022)} \cite{chollet2022training} & Public speaking simulation with virtual audience; after-action report  & Implicit audience behavioural feedback (nods, gaze, postures) \& after-action  feedback interface: graphs,  comments  & Facial expressions, pauses, gaze & \textbf{Evaluation: 57 users}, improved confidence, eye contact, flow of speech, only when after-action report contained personalised feedback. \\ \\

\textbf{LISSA} \cite{Ali2015} & Conversational agent for social and language skills & Verbal and visual interaction cues & Eye contact, smile, volume, body movement & \textbf{47 users}. 
Improved head nods, effective gesturing and eye contact. \\ \\

\textbf{MACH} \cite{Hoque2013} & Simulated interview trainer with embodied agent & Post-session feedback and replay & Speech (tempo, fillers), prosody (loudness, intonation, pauses), face (head nods, smile)
%Gaze, posture, pitch, tempo 
& \textbf{90 users}. Improved interview skills and self-presentation. \\ \\

 \textbf{PASCAL} \cite{NiebuhrPascal2017} & Prosodic coaching for persuasive speaking & Curve-based visualisation of prosodic features for acoustic feedback & Pitch and loudness  & \textbf{45 users}. Boosted persuasive impact and vocal charisma. \\ \\

\textbf{Rhema} \cite{Tanveer2015} & Wearable real-time speech coach using Google Glass & On-glass visual feedback on vocal features & Speaking rate and volume & \textbf{30 users}. Compared evaluation of feedback schemes. Effective in increasing awareness of vocal behaviour. \\ \\

\textbf{RocSpeak} \cite{Zhao2017} & Web-based system with automated and crowd feedback & Combined video analysis and peer input & Audio (loudness, pitch), visual (smile, movement), speech (unique words count) & \textbf{56 users}. Improvement of speaking performance, increased confidence.\\ \\

\textbf{Web Pitcher} \cite{NiebuhrWebPitcher2021} & Web-based pitch training with prosodic feedback & Two visual feedback modes: curve and  traffic-light modes. & Pitch, tempo, pauses & \textbf{60 users}. Increased pitch range, more frequent silent pauses, and a slower tempo.% and expressive modulation.
\\

\hline

\end{tabular}

\label{tab:users_research_tools}
\end{table*}

\begin{table*}[t]
\centering
\caption{Overview of academic systems for public speaking training designed and/or evaluated with experts.}
\footnotesize
%\begin{tabular}{p{1.5cm} p{2.8cm} p{3.3cm} p{2.8cm} p{4cm}}
\begin{tabular}{p{1.6cm} p{2.3cm} p{3.2cm} p{3.1cm} p{4.2cm}}
\\
\hline
\textbf{Tool} & \textbf{Description} & \textbf{Feedback} & \textbf{Cues} & \textbf{Design and/or Evaluation} \\
\hline
\\

\textbf{Automated Skills Trainer} \cite{tanaka2023validation} & Multimodal coaching tool, 4 role-play scenarios & Performance metrics presented on a radar chart, comments, suggestions & Voice (vocal variation, clarity, fluency), facial (gaze, facial expressions), body orientation, and social validity& \textbf{Evaluation: 18 users, 2 experts}. Improved delivery and confidence. \\ \\

\textbf{Cicero (2015)} \cite{chollet2015exploring} & Virtual audience simulation for speaking practice & Implicit feedback based on virtual audience attitudes & Prosody, posture, gaze & \textbf{Evaluation: 47 users \& 3 experts}, improved overall public speaking performance as rated by experts. \\  \\

\textbf{EmoCo} \cite{Zeng2019} & Interactive visual analysis of emotion coherence & Video view; multimodal emotion channel coherence view, word/sentence clustering view, etc  & Face, audio, and text features, multi-modal and multi-level emotion alignment & \textbf{Design: 2 experts}, \textbf{Evaluation: 2 experts}  (interviews);  efficient and insightful analysis of emotions. \\ \\

\textbf{GestureLens} \cite{Zeng2022} & Real-time gesture analysis tool  & Gesture exploration view, and speech-gesture relation view  & Metaphoric and beat gestures & \textbf{Design: 2 experts},           \textbf{Evaluation: 2 experts} (interviews); evidence for improving gesture usage.  \\ \\

 \textbf{Presentation Trainer} \cite{schneider2017presentation} & Rule-based real-time  non-verbal behaviour  feedback & Recognition of a set of "ineffective practices" (e.g. crossing arms, insufficient pauses, etc) & Nonverbal cues; voice volume, posture, use
of pauses and gestures& \textbf{Evaluation: 10 experts}; Identification of (in)effective non-verbal communication practices. \\ \\

\textbf{RAP}  \cite{ochoa2020controlled} & Feedback on slides and non-verbal behaviour  & Report on overall presentation quality with global, detailed scores, pre-generated advice & Audio (filled pauses, volume), gaze, body posture, slides quality & \textbf{Evaluation: 180 users, 10 experts}. Improvement in gaze behaviour, reduction in filled pauses. Significant, albeit small, expert-rated improvement. \\
\\

\textbf{SpeechMirror} \cite{Huang2024} & Visual analytics dashboard for online speech self-reflection & Personalised visualizations, data-enhanced video playback & Facial cues (valence, arousal), gaze, body gesture energy, stage usage, voice (loudness,  pitch, pace), script & \textbf{Design: 2 users, 4 experts}, \textbf{ Evaluation: 4 users, 4 experts}; Actionable insights for improving behaviour. \\ \\

\textbf{VoiceCoach} \cite{Wang2020} & Interactive system to explore and practice voice modulation skills & Interactive visual analytics: recommendations view, voice technique view, and practice view & Speech: Pause, volume, pitch, speed  & \textbf{Design; 18 users \& 2 experts}, \textbf{Evaluation: 24 users} 
\& \textbf{4 experts} Positive impact and effectiveness of the system in helping novice speakers with the training of voice modulation skills.  \\ 
 \\

\hline

\label{tab:experts_research_tools}

\end{tabular}

\end{table*}

Recent years show increasing interest in AI-driven public speaking training tools, spanning
academic research and commercial innovation. This section reviews systems that 1) leverage AI
for automatic speaker behaviour analysis and 2) have been designed/evaluated with
end-users/experts.

% Recent years have witnessed an increasing interest in developing AI-driven tools aimed at supporting the training on public speaking skills. A wide range of initiatives span both academic research and commercial innovation in industry. This section reviews  public speaking training systems developed as part of academic research prototypes as well as commercial applications. Firstly, we provide an extensive analysis of research prototypes with a particular focus on systems which 1) leverage Artificial Intelligence for the automatic analysis of the speaker's behaviour and 2) that have been designed and/or evaluated with end-users and/or experts. Secondly, we present a selection of commercial tools which propose automated practice solutions for the training of public speaking skills.

\subsection{Academic research on public speaking training systems}
\label{background_academic}
The past decade has seen a surge in academic AI-powered systems for public speaking. These
prototypes integrate multimodal social signal analysis for automated feedback. Tables~\ref{tab:experts_research_tools} and~\ref{tab:users_research_tools} summarise systems that have been evaluated in user studies, based on: 1) assessed speaker (non-)verbal cues, 2) feedback provided by the system, and 3) evaluation results with end-users or experts.

% The past decade has witnessed a surge in the development of academic AI-powered systems designed to support and enhance public speaking skills. These research prototypes integrate multimodal social signal analysis, capturing verbal and nonverbal cues 
% to provide automated or semi-automated feedback. Table \ref{tab:experts_research_tools} and Table \ref{tab:users_research_tools} summarizes a review of the systems discussed in this section based on; 1) verbal and non-verbal behavioural cues used for the automatic assessment, 2) the feedback returned by the tool, 3) the tool's evaluation conducted with end-users only (resp. with end-users and experts) in a public speaking training setting. 

Numerous academic systems have emerged with distinct focuses. GestureLens \cite{Zeng2022} offers visual overlays of gestural dynamics. VoiceCoach \cite{Wang2020} and Web Pitcher \cite{NiebuhrWebPitcher2021} focus on prosodic features. Presentation Trainer \cite{schneider2017presentation} %and AwareMe \cite{Bubel2016} rely
 relies on body posture and movement cues. The feedback provided by these systems may focus speaking rate, vocal clarity, filler words, gesture size, gaze focus, posture, and more. Some tools adopt unique perspectives:  EmoCo \cite{Zeng2019}  emphasises emotional expressiveness; RocSpeak \cite{Zhao2017} combines automated behavioural analysis with crowdsourced peer feedback; AutoManner \cite{Tanveer2016}, identifies behavioural mannerisms (e.g. repetitive motions). These systems differ in the type (visualisations, quantifications), modality (visuals, text, wearable haptics), and temporality (real-time, post-training) of feedback. Other tools like LISSA \cite{Ali2015},  MACH \cite{Hoque2013} and Cicero \cite{chollet2022training} simulate social situations using embodied agents capable of (non-)verbal behaviour expression.

% The above systems are based on off-line feedback. Other tools, such as Rhema \cite{Tanveer2015} and LISSA \cite{Ali2015}, incorporate real-time wearable interfaces or virtual agents to deliver feedback during live or simulated interactions. MACH \cite{Hoque2013} and Cicero \cite{chollet2022training} focus on the simulation of social situations, such as job interviews and public speaking, through the use of embodied conversational agents, such as virtual job recruiters or virtual audiences. The feedback modalities covered by these systems vary widely between offline and real-time feedback. Offline feedback methods often rely on visual graphs based on different data visualisation techniques. Real-time displays imply the use of wearable-specific devices such as a wristband for haptic feedback in the AwareMe tool \cite{Bubel2016} and head-mounted displays for real-time visual displays in \cite{damian2015augmenting}. 

% The evaluation of these systems is essential not only to validate their usability and effectiveness but also to uncover shortcomings and opportunities for improvement. 
Most studies incorporate user evaluations with subjects of diverse public speaking proficiency. Sample sizes range from small pilot groups (e.g., 10 to 20 users)  \cite{Tanveer2016,Zeng2022} to more extensive evaluations (e.g., 40+ users), as in RocSpeak \cite{Zhao2017} or Automated Skills Trainer \cite{Tanaka2015} and Cicero \cite{chollet2022training}), with the randomised evaluation of the RAP system by Ochoa et al. in an educational setting being the largest and most ecologically valid \cite{ochoa2020controlled}. 
% These studies tend to measure improvement in speaker confidence, expressiveness, and nonverbal coordination, often through pre/post-session third-party assessments by experts or laypeople, or by subjective self-assessments. 
Evaluations assess improvements in non-verbal behaviours as well as  in varied ratings such as confidence, expressiveness, or non-verbal coordination, often via pre/post-training third-party ratings or self-assessments. For instance, VoiceCoach led to significant improvements in vocal dynamics \cite{Wang2020}. Rhema helped regulate pacing and reduce filler words \cite{damian2015augmenting}. GestureLens increased  gestural awareness \cite{Zeng2022}, and RocSpeak led to increased self-confidence and performance ratings \cite{Zhao2017}. Such findings highlight the role of automated feedback in building public speaking skills.

While promising, academic prototypes often evaluated user satisfaction, engagement, or perceived improvement of behavioural metrics after isolated sessions. This overlooks holistic trainee journey and  how training activities integrate within it. Though valuable, these evaluations are insufficient to fully understand systems' pedagogical efficacy and long-term benefits for learners.

% While the academic prototypes for public speaking training described above have demonstrated promising capabilities, we observe that their evaluation strategies often center around user satisfaction or engagement ratings, or perceived improvement of specific behavioural metrics after an isolated public speaking practice session, overlooking the rest of a public speaking trainee's journey and how such training activities should integrate in it. This approach, though valuable, is insufficient to fully understand the pedagogical efficacy and potential of such tools to deliver long-term, meaningful benefits for learners.

\begin{table*}[t]
\caption{Comparative analysis of selected commercial public speaking training systems.}
\label{tab:tool_analysis}

\small

\begin{tabular*}{\textwidth}{@{\extracolsep{\fill}}p{1.7cm}p{4.4cm}p{4.4cm}p{4.4cm}@{}}
\toprule
\textbf{Tool} & \textbf{Description} & \textbf{Feedback} & \textbf{Cues} \\
\midrule

\parbox[t]{1.7cm}{
\textbf{Polymnia}\\
Web platform\\
Lang.: {\scriptsize \selectfont FR, EN, DE, ES, IT, PT} }  
& Browser-based training tool; marketed to independent users or as an e-learning professional service. Independent rehearsal feedback, and 10-step training program. 
& Real-time behavioural feedback. Post-training: dashboard, behavioural metrics, lexical/structural analysis. Speech dynamics visualization. Recommendations, e.g. word choice, pauses. & Facial, bodily cues (e.g. posture, smile, gesture type). Speech: prosody (pace, silences), lexical content (e.g., word frequency). Detailed report of speech metrics. \\  \\

\parbox[t]{1.7cm}{\textbf{VocaCoach}\\ Web platform \\ Lang.: {\scriptsize\selectfont FR, EN}} 
& For independent users or professional teams. Users upload or record video performances; system provides immediate feedback, and analysis of progression.  
& Scores (1–5) on 5 main speech criteria. Additional cues (e.g., silences, repetitions) analysed without scoring. Trend plots; behaviour-specific examples.
& Speaking pace (words per second), hesitations (filler words count), rhythm (sentence length, pause detection), articulation ("acoustic clarity"), tone (pitch modulation).  \\ \\

\parbox[t]{1.7cm}{\textbf{Speaker Coach}\\
MS Teams plug-in. Lang.:{\scriptsize\selectfont EN}} & For professionals using Microsoft Teams; real-time speech analysis during live meetings (no recording saved); private feedback report. & On-screen live tips; short post-meeting summary with recommendations. & Filler words, pace (speech rate), intonation (pitch variation), inclusiveness (speaking turn duration analysis), monologues, repetitions. \\  \\

\parbox[t]{1.7cm}{\textbf{Poised}\\
Desktop, web app \\
Lang.:{\scriptsize\selectfont EN}} & Analysis of video meetings (e.g., MS Teams, Zoom). Real-time and post-meeting feedback. Blog, TED talk analyses,  speech templates. AI role-play via QA bot and  assistant. & Post-meeting report. Feedback is based on user-defined goals (e.g., inspire, inform, lead). Highlights strengths and areas to improve with examples of user phrasing. & Pacing (e.g., speech rate), disfluencies (filler words), structure (e.g., topic coverage), jargon (e.g., term clarity), conciseness (redundancies), etc. \\ \\

\parbox[t]{1.7cm}{\textbf{Orai}\\
Mobile app (iOS/Android) \\
Lang.:{\scriptsize\selectfont EN}} 
& General public, young professionals. Short video-based lessons. Structured exercises and applied tasks, e.g. read aloud, improvise speech, quizzes. 
& Feedback: scores, visualisations. Score for overall performance and for Energy, Conciseness, Confidence, Pace. Highlighted transcripts, storytelling evaluation activities, examples.
& Words per second (pace), pitch, volume range (energy), filler words and long pauses (confidence), repetitions, passive voice, avg. sentence length (conciseness). \\

\\

\parbox[t]{1.7cm}{\textbf{Yoodli}\\
Web platform \\
 Lang.:{\scriptsize\selectfont EN, ES, FR, PT, IT, CH, JAP, GR, HIN, KOR, DU, PL}} & Marketed for professional teams. Asynchronous video analysis and real-time meeting feedback. Structured courses, roleplay simulations, interview training, impromptu speech practice.  &  Overall performance score (\%), detailed AI-generated coaching on strengths, growth areas, conciseness, demeanour. Visual analytics on specific cues. In roleplays, breakdown per response. & Word per minute, pauses (pace); smiles, eye contact duration; filler words; weak/repetitive word use sentence length; question count; talk time (in meetings). \\ \\

\parbox[t]{1.7cm}{\textbf{VirtualOrator}\\
VR application \\
Lang.:{\scriptsize\selectfont  R, GR, EN} }& Virtual Reality simulations with configurable venue and audience size. Designed for overcoming public speaking anxiety, skills training, and preparation.  & Post-session feedback includes presentation timing, duration, eye contact, and a summary of simulated audience reactions. & Presentation length, speech timing, eye contact tracking. Possibly behavioural simulation metrics, but no precise cues listed (e.g., no word-per-second metrics). \\
\\

\parbox[t]{1.7cm}{\textbf{VirtualSpeech}\\
Web + VR  \\
 Lang.:{\scriptsize\selectfont 14}} & Online platform with VR and AI-based scenarios: e.g., interviews, performance reviews, mentoring, etc. Custom environments and scenarios.
& AI roleplay: overall score, transcript, eye contact, body language, filler words, pace, loudness, keywords, "listenability". & Pace (e.g. words per second), filler words, loudness, body posture, eye contact, keyword frequency. In VR: Eye contact, body language.\\

\bottomrule

\end{tabular*}

\end{table*}

\subsection{Commercial applications for public speaking training}
\label{background_commercial}

% In addition to the academic research on the development and the evaluation of public speaking training prototypes, a growing number of commercial applications have emerged for public speaking training. 
% In this section, we briefly present a set of commercial public speaking support tools through a series of searches realised by the authors on general search engines throughout the summer of 2023. We provide a detailed description of these systems in Table \ref{tab:tool_analysis}. This selection of tools was chosen not only for their technological diversity 
% but also for their visibility, adoption, and relevance to both French-speaking and international users. Together, they reflect a representative sample of the current landscape of automated public speaking coaching solutions and illustrate how recent commercial solutions are framing technologically-mediated public speaking training paradigms.

Beyond academic research, numerous commercial public speaking applications have emerged. This section presents a selection of these tools, identified in  web searches conducted in summer 2023, and chosen  for their technological diversity, visibility, and  adoption (see Table~\ref{tab:tool_analysis}). They represent the current landscape of automated public speaking coaching solutions and illustrate recent technologically-mediated training paradigms.

Microsoft Speaker Coach, integrated into PowerPoint, offers feedback on vocal (pace, filler words, intonation, repetitive language)\footnote{\label{footnote-speakercoach}\url{support.microsoft.com/en-us/office/rehearse-your-slide-show-with-speaker-coach-cd7fc941-5c3b-498c-a225-83ef3f64f07b}}. VocaCoach\footnote{\url{vocacoach.fr}} also focuses on vocal analysis (pitch variation, volume, speech rate, pause frequency). PolymnIA\footnote{\url{polymnia-france.com}} stands out with multimodal feedback, analysing prosody, gestures, postures, and facial expressions. Tools like Poised\footnote{\url{poised.com}}, Orai\footnote{\url{orai.com}}, and Yoodli\footnote{\url{yoodli.ai}}  leverage AI to provide personalised speech coaching through mobile/desktop applications. VirtualOrator\footnote{\url{virtualorator.com}} and VirtualSpeech\footnote{\url{virtualspeech.com}} propose  VR environments to simulate public speaking scenarios.

% In Table \ref{tab:tool_analysis} we present an analysis of the main features of these commercial public speaking training tools. We describe the behavioural cues analysed by the system as well as the visual feedback provided by the system in order to help users identify areas of improvement for their communication skills.  We discuss these systems further in section \ref{system_analysis_interview} where we describe their main features according to the analysis of semi-structured interviews conducted in our study (see section \ref{study1-itw}).

Commercial tools often build upon similar behavioural cues to research prototypes, while expanding their application scope through multilingual support and user-friendly design. However, they do not disclose how behavioural cues are aggregated into feedback, nor do we have access to evaluations of their effectiveness. 

\subsection{The role of experts in the design and evaluation of public speaking training systems}
\label{background_experts}

Expert evaluations offer critical insights complementing user-based assessments, particularly for assessing subtle performances. Some academic research \cite{schneider2017presentation,Zeng2019,Wang2020,Zeng2022} incorporated public speaking experts during system design or evaluation. For example, VoiceCoach \cite{Wang2020} showed improved vocal performance assessed by two experts. Tables~\ref{tab:experts_research_tools} and~\ref{tab:users_research_tools} overview research systems with/without expert involvement. Experts often evaluate systems, providing trainee improvement annotations; they are sometimes advisors in early design phases, though this is uncommon.

However, while these research projects offer insights into automated public speaking feedback design and benefits, no study has evaluated commercial public speaking training systems. With such systems widely available and poised to disrupt public speaking training, critical evaluation is crucial. What opportunities do these systems bring to augment traditional coach-based frameworks? What challenges face learners relying solely on these systems? How should next-generation AI-based systems leverage human coaches and AI-feedback strengths for optimal learner outcomes?

We believe experts are essential to guide system evaluation, refinement, and broader adoption in educational/professional contexts. To our knowledge, our work is the first to explore public speaking training experts' opinions on commercial AI-driven public speaking training systems' reliability and design.

% However, while these research projects give us general insights about the design and potential benefits of automated public speaking feedback, no study has evaluated commercial public speaking training systems. Now that such systems are largely available and are poised to disrupt the ecosystem of public speaking training, there is a crucial need to critically evaluate them, and to consider the place of such automated feedback systems in public speaking trainees' journeys. What opportunities do they bring to augment traditional coach-based public speaking training frameworks? What challenges are faced by learners relying solely on these systems? How should the next generation of AI-based public speaking training systems be designed to leverage human coaches and AI-feedback systems respective strengths to deliver the best possible outcomes for learners? 

% We believe experts will be essential to guide the evaluation and refinement of these systems and to support their broader adoption in educational and professional contexts. To the best of our knowledge, our work constitutes the first attempt to explore the opinion of public speaking training experts on the reliability and the design of commercial AI-driven public speaking training systems. 

\section{Methodology}\label{methodology}

This research includes two studies involving public speaking coaches. The first explores current training practices and learner needs through semi-structured interviews and design workshops. The second investigates how existing commercial systems may be integrated into training, focusing on expert evaluations of their usefulness and limitations.

% In this research project, we set out with a first objective of understanding modern public speaking training practice through the perspectives of public speaking coaches. We sought to ground the project by engaging in participatory design research, involving public speaking experts in reviewing the current landscape of public speaking training applications, identifying the needs of public speaking trainees, and designing for the next generation of training tools. Our methodology involved in-depth, semi-structured interviews followed by design workshops with public speaking training coaches.

\textbf{Study 1 (Section \ref{study1-itw}):} To ground our research in real-world practice, we conducted interviews with public speaking experts. The aim was to leverage their expertise to anchor our understanding of how public speaking is conceptualised, evaluated, and taught, rather than relying on theoretical assumptions or existing technological frameworks. This grounded understanding informed
the rest of our project, especially Study 2.

% First and foremost, we decided to organise a series of interviews with public speaking experts to ground our understanding of public speaking, and of the best practices to teach it, in the context of their practice. The goal of this first study was to let experts' in-depth understanding of public speaking guide our research instead of starting with assumptions grounded in theory or dictated by technological choices made in existing public speaking training systems. In a nutshell, we started with general interview with public speaking training experts to understand how they conceptualise public speaking as an activity, how they evaluate public speaking quality, how they approach its training, and so on. This grounded understanding of public speaking training was then used to organise the rest of our research project, and in particular the second study.

\textbf{Study 2 (Section \ref{study2-wrskp}):} we conducted design workshops to gather experts insights on  existing commercial or prospective computer-assisted public speaking training systems, to identify recommendations and guidelines for designing more effective tools aligned with real public speaking practice. 

\section{Study 1: Individual Interviews with Public Speaking Coaches}\label{study1-itw}

This section introduces our first study: interviews with public speaking experts on their practice,
trainees, and training views, grounding our research in domain expertise.

\subsection{Participants}

We recruited 16 expert public speaking coaches via LinkedIn and our professional networks for voluntary, uncompensated interviews.Seven (6F, 1M) were in-person, and nine (6F, 3M)  via phone/online, using the same semi-structured interview format (described in Section \ref{interview_protocol}.  Our heterogeneous sample reflects coaches with diverse client ranges, including work with individuals/groups, companies, adults/adolescents, or exclusively women.

\subsection{Semi-structured Interview Protocol}
\label{interview_protocol}
We prepared a semi-structured interview guide covering several main themes.

\textbf{Theme 1: Trainees' profile and motivation and training sessions' context} focused on trainee
profiles/motivations and training contexts:

\begin{itemize}
    \item Trainee identity/profile (individuals, business clients, etc.)
    \item Training session context (individual/group sessions)
    \item Trainee motivations (reasons for engaging with public speaking training)
\end{itemize}

\textbf{Theme 2: public speaking performance evaluation} clarified how coaches evaluate public speaking performances:

\begin{itemize}
    \item Focus (or absence thereof) on particular sub-skills or specialities
    \item Categories/taxonomies of  different speech types 
    \item Main ingredients of good public performance and high-level assessment factors
    \item Measurement methods and systematic tools used
\end{itemize}

\textbf{Theme 3: coaching activities} focused on training practices:  

\begin{itemize}
    \item How trainers guide learners through their journey (main coaching steps)
    \item Dealing with public speaking anxiety (given its prevalence, \cite{furmark1999social})
    \item Role of practice, and what practical activities they use
    \item Consolidating trainees' learning over time
    \item The place of technological systems to improve practice and their imagined functions
    \item Awareness or use of existing technological platforms for public speaking training
    \item Knowledge of public speaking certifications
\end{itemize}

\subsection{Interviews Analysis and Key Results}

We used inductive thematic analysis to analyse the data. After transcribing interviews, we familiarized ourselves with the data and generated themes. Themes were then iteratively redefined, discussed, combined, and finalized, by two of the paper authors.

\subsubsection{Public speaking trainees’ needs } A recurring comment was that public speaking trainees have ``\textit{ideas to express, but not always the skills}.'' Most clients have specific goals and opportunities driving their desire to improve, rather than a general aim to become better speakers. They face challenging professional situations or have short-term future public speaking opportunities. Identifying this specific need and context is typically the coaches' first step.

% A comment that came up regularly in the interviews was that public speaking trainees have '' \textit{ideas to express, but not always the skills}''. Most of our participants' clients come with specific goals and public speaking opportunities in mind that spur their desire to improve. That is, public speaking trainees typically have a specific need or set of circumstances which drives their desire to improve their public speaking skills. They do not reach out to a public speaking coach with a general goal to become better speakers: they are rather faced with particular professional speaking situations which they find challenging, or have a specific professional goal involving public speaking in the future. Identifying and understanding this specific client need and its context is typically the first step of coaches when they take on a new client.

However, trainees present various challenges. Some feel illegitimate speaking out. All 16 coaches noted a large proportion deal with public speaking anxiety, desiring to overcome fear or feel more at ease. Coaches first build self-confidence, which brings natural improvements in performance; after initial confidence is established, activities building oratory skills further strengthen confidence. Only a minority of clients start with high self-confidence, seeking specific efficacy improvements.

% However, trainees come with various types of challenges which they see as obstacles to achieve their public speaking goals. Some trainees reach out to coaches saying they do not feel legitimate in speaking out. All 16 coaches mentioned that a large proportion of their trainees deals with a degree of public speaking anxiety, expressing a desire to no longer be afraid or to feel more at ease when speaking in public. The technical aspects of improving as a speaker generally come afterwards this first issue of improving self-confidence in public speaking, however, coaches note that improving their clients’ oratory skills generally further strengthens their self-confidence. Overall, it's only a minority of clients which come to coaches with an already high level of self-confidence, with a specific desire to improve their public speaking efficacy.

Coaches overwhelmingly report that trainees initially have a very critical view of their own skills, believing audiences perceive them negatively. This negative bias is prevalent; many clients don't actually exhibit terrible skills but need cognitive restructuring to reframe these negative self-perceptions.

% Furthermore, coaches overwhelmingly report that trainees come first with a very critical view regarding their own speaking skills, the majority holding a strong belief that audiences perceive their speaking performance negatively. This negative bias towards their own behaviour seems very prevalent, coaches often highlighting that many such clients do not actually exhibit particularly terrible speaking skills, but require some degree of cognitive restructuring to reframe these negative self-perceptions. 

\subsubsection{Dimensions of speech quality}
Beyond trainee needs, our interviews identified public speaking quality dimensions. We found
three major and three secondary dimensions coaches use for evaluation:

% Another major aspect of our interviews, in addition to the needs of the trainees, is related to the identification of the dimensions of public speaking quality. We identified three major and three secondary dimensions through which coaches use as a framework to evaluate the quality of public speeches. The main dimensions are (1) \textbf{content}, (2) \textbf{form}, and (3) \textbf{emotions} induced in the audience.

\begin{enumerate}
    \item \textbf{Content:} Focuses on word choice, speech structure, and argument
elaboration. Coaches recommend adapting messages to the targeted audience.
    % The content is linked to the specific words and structure of the speech and how it is elaborated. It concerns what arguments and information speakers choose to include or not to include to deliver their message, and their choice of words. Here, trainers’ main recommendation is to encourage speakers to identify their targeted audience to correctly adapt and structure their message to it.
    \item \textbf{Form}:  Encompasses non-verbal social cues, including visual (gestures,
facial expressions, posture, space use) and vocal dimensions (tone, intonation, speaking
rate, intensity, articulation, pronunciation).
    % The form of a public speech corresponds to a wide variety of non-verbal social cues. This includes visual, bodily behaviours: gestures, facial expressions, smiles, gazes, postures, the use of space, etc. The vocal dimension of public speaking also falls in this scope: tone of voice, intonation, speaking rate, intensity of voice, articulation, pronunciation, and in general all aspects related to speech delivery.
    \item \textbf{Emotions:} Represents the intended outcome, linked to the speaker's goals (to
inform, persuade, inspire, their audience).
    % Most interestingly, the emotion induced in the audience is, in a sense, the intended outcome of public speaking activity. This is highly related to a speaker’s main goals in undertaking public speaking, and whether it is to inform, persuade, inspire, an audience.
\end{enumerate}

Three additional  secondary themes emerged from the interviews: (4) effective use of presentation media, (5) management of interactions with the audience, and (6) what several coaches called ``little extras'' (\textit{petit plus})—ill-defined, highly personal, subjective qualities that distinguish exceptional speakers.
% Three secondary dimensions emerged from the interviews with coaches: they concern (4) the appropriate use of presentation media, (5) the management of interactions with the audience, and (6) what several coaches called the ``little extras'' (\textit{petit plus}), i.e. an ill-defined category of highly personal, subjective elements which allow particularly remarkable speakers to stand out, differentiating competent, well-trained speakers from truly exceptional speakers.

%\subsubsection{Three moments of public speaking}
\subsubsection{Pillars of coaching activities}
After synthesising interviews and readings, we group public speaking coaches’ activities, tools and exercises around three pillars, consisting in different temporalities and activities of public speaking activity.

\textbf{(1)	Preparation -} This pillar involves actions taken well in advance for a public speaking event (e.g., conference, meeting). It is crucial to define objectives and select content to achieve them. It encompasses setting goals and scoping content: what will be said, and why. Anticipating the material conditions of the speech (e.g. at a pulpit, standing or sitting, etc) is also crucial.
% This pillar corresponds to all actions that should be taken to prepare for a public speaking event well in advance, particularly in the case of a conference, presentation or running a meeting. A key theme here is taking the time to define their objectives for a particular presentation, and choosing the elements of content to include and not to include, in order to allow us to achieve this objective. In a nutshell, this involves the actions of preparing one’s goals, and the scope of a presentation’s content: what will be said and why. Another aspect is also anticipating the material conditions for speaking to prepare accordingly.
\begin{quote}
    ''There is no good speaker who is not a prepared speaker.'' P3
\end{quote}

\textbf{(2)	Stress management -} This pillar covers methods for managing stress before speaking, either weeks or days prior, or minutes before the speech, to handle performance anxiety. Techniques often include breathing, relaxation, visualization, or cognitive restructuring.
% The second pillar concerns all the methods that allows a speaker to manage stress before speaking. In terms of temporality, these happen either well before speaking (i.e. in the weeks or days preceding the activity), or in the hours or minutes before the speech itself to manage performance anxiety. These  often involve breathing methods, but relaxation, visualization, or cognitive restructuring techniques are also taught by some trainers in our sample.
\begin{quote}
    ''For 75\% of people who are stressed, having methods and techniques [to manage stress] is extremely reassuring. Especially during bad days, you are able [thanks to these techniques] to hang on to [your technique].'' P2
\end{quote}

\textbf{(3)	Delivery -} This final pillar concerns managing all behavioral aspects of delivering a speech to achieve a great public speaking performance.\begin{quote}
    ''It’s really a combination that will make us effective. A combination of rhythm, tone, posture, voice, look. If the gaze is shifty, it sends out lots of indicators of lack of confidence which will cause listeners to doubt what is being said, unconsciously.'' P1
\end{quote}

 \subsubsection{Features of speech delivery}
Coaches describe public speaking delivery along three modalities: 1) \textbf{vocal}, 2) \textbf{verbal}, and 3) \textbf{visual}. They suggest behaviour-specific exercises tailored to each client’s needs. Below is a non-exhaustive list of delivery elements mentioned by several coaches.
% Coaches seem to separate public speaking delivery along three dimensions: 1) \textbf{vocal}, 2) \textbf{verbal} and 3) \textbf{visual}. Their training activities regarding delivery included a range of behaviour-specific exercises that their suggest to clients according to their specific needs. Below is a non-exhaustive list of elements of delivery mentioned by several coaches.
\begin{itemize}
    \item \textbf{Verbal behaviour}: Formulate simple sentences; Speak slowly, add meaningful silences; Use a precise level of language adapted to the audience; Use metaphors or analogies to explain new and complex ideas; Eliminate language tics and mannerisms.
    \item \textbf{Visual behaviour}: Avoid self-contact and non-verbal tics; Use the space: walk, but do not trample; Use gestures to explain, and to maintain attention; Look at your audience.
    \item \textbf{Vocal behaviour}: Project your voice; Take the time to breathe; Articulate; Modulate your voice according to key ideas and sentences, on which intonation will be more pronounced.
\end{itemize}

\subsection{System-level analysis according to semi-structured interviews themes}
\label{system_analysis_interview}
In Table ~\ref{tab:tool_analysis} and in Section~\ref{background_commercial}  we summarise the \textbf{description}, \textbf{feedback} and the main \textbf{cues} analysed by the several internationally used public speaking training systems. Below, we contrast  system features  to the \textbf{three major themes} evoked by coaches during the Study 1 interviews.  

\noindent \textbf{Theme 1: Trainees' profile, context, and motivation. }  Most systems target general users without assuming prior training. VocaCoach, Orai, and Polymnia are designed for independent learners without institutional onboarding or coach supervision. Speaker Coach or Poised integrate into professional contexts (meetings, team communication). These platforms operate asynchronously, supporting individual use. While tools imply learner-centred logic (improving fluency, reducing filler words, increasing vocal dynamism), they rarely explicitly inquire into learners' training goals or presentation contexts.

\noindent \textbf{Theme 2: Public speaking performance evaluation. } Systems differ in analysed features. VocaCoach focuses on vocal criteria (pace, tone, articulation, rhythm, hesitation) with numeric scores, while Polymnia or Yoodli combine verbal, vocal, and visual analysis. Several tools summarise low-level cues (pitch variation, smile frequency) into broader dimensions (confidence, clarity, conciseness). However, mechanisms computing these higher-order criteria remain opaque: it's unclear whether they're based on rhetoric/psychology literature, coach feedback, or empirical heuristics/machine-learning based models. No application provides information on the underlying rules, thresholds, or systems enabling them.

Some contextual adaptability exists. Polymnia lets users select audience type, relationship, and speech goal (inform/persuade), though how this impacts feedback is unclear. Poised offers goal-tailored feedback ("inspire"/"lead"), but again how this impacts feedback logic isn't explained. 

\noindent \textbf{Theme 3: Training pedagogical activities. } Some tools propose training plans. Orai provides structured modules on accent reduction, storytelling, interviews, while Yoodli includes courses on public speaking basics and strategic delivery. These combine self-practice with educational components (expert tips, pedagogical videos). Nevertheless, no platform adopts long-term coaching structures with systematic learner progression tracking and coach support. Several systems allow accessing past recordings/feedback—offering basic progress review. However, they don't provide progression visualizations or integrate past performances into future training advice.

No platform includes public speaking anxiety reduction exercises (breathing techniques, visualization). This is a notable gap, as stress management is critical in coaching practice. Tools focus mainly on speech delivery, with limited attention to preparatory or psychological dimensions.

\noindent \textbf{Overview on contrasting commercial framework within expert-informed grounding.}  Most existing tools target individual, general users, with limited customisation to specific goals or performance criteria. Evaluation frameworks typically rely on observable features and aggregated scores, yet they often lack transparency and theoretical grounding connecting these features to the feedback provided. Adaptation to different languages and cultural contexts remains unclear, particularly in how language influences scoring mechanisms. The role of AI in generating feedback is rarely transparent. While some tools, such as Polymnia and Poised, offer contextual customisation, its impact on feedback remains poorly understood. Integration with structured coaching or pedagogical progression is minimal—only Orai and Polymnia include curriculum elements, but tracking of learner progress is limited. Although some systems include educational content, their quality and depth vary considerably. Notably, none of the reviewed systems incorporates stress management or anxiety-reduction techniques, such as breathing exercises, which are common in traditional coaching.

% \vspace{0.3cm}
% \noindent\textbf{Key insights:}
% \begin{itemize}
%     \item Most tools target individual, general users; Customisation to the user's goals and targeted criteria is limited.
%     \item Evaluation frameworks are based on observable features and aggregated scores, but lack transparency and literature grounding connecting observable features and recommendations.
%     \item Lack of adaptation to the context of different languages and cultures; Specifically, it is not clear how scoring based on the features is affected by the change of language. 
%     \item Lack of transparency in the role of AI in the feedback. 
%     \item A few systems offer contextual customization (e.g., Polymnia, Poised), though it's unclear how it shapes feedback.
%     \item Learning is not coach-integrated; curriculum elements exist in Orai and Polymnia but progression tracking remains limited.
%     \item Educational content is present in some systems, but varies in quality and scope.
%     \item Stress management and anxiety-reduction exercises (e.g., breathing) are absent from all analysed systems.
% \end{itemize}

\section{Study 2 - Design Workshops}
\label{study2-wrskp}

Initial interviews (Section~\ref{study1-itw}) provided a better understanding of public speaking coaching and coaches' mindsets. This also helped establish relationships with coaches, whom we re-contacted for Study 2 on our core research goal: exploring expert opinions on the usefulness, usability, and design of commercial computer-based public speaking training systems. To explore these goals, we organized follow-up workshops with a subset of coaches. We presented existing public speaking training systems, collected opinions, and conducted speculative design activities to identify opportunities. Main questions explored:

% The initial interviews presented in section \ref{study1-itw} provided us with a better understanding of the global context of public speaking coaching activities and of the frame of mind of coaches regarding public speaking in general. It also allowed us to establish a relationship with a set of public speaking coaches that we then contacted once more to study our main research goals: investigating expert opinions of commercial computer-based public speaking training systems in terms of their usefulness, usability, and potential design improvements.

% To explore these research goals, we organized follow-up workshops with a sub-set of the public speaking coaches in which we presented them with several existing commercial and/or research public speaking training applications, collected their opinions, and undertook speculative design activities to identify design opportunities. Below are the main questions explored in these workshops:

\begin{itemize}
    \item Is a technological public speaking training system truly useful? In what formats, contexts,
and for what tasks?
    \item Do trainers believe such systems allow independent evolution, without a personal
coach?
    \item Can the system integrate into coaches' training practice? If so, how?
    \item What key features would complement coaches' practice, benefitting them and trainees alike?
\end{itemize}

\subsection{Participants}
We recruited a subset of public speaking experts from our initial interview study (Section \ref{study1-itw}), aiming for complementary specialities. Before workshops, experts completed an online questionnaire on their generalist (\textit{resp.} specialist) status and their central (\textit{resp.} unaddressed) topics (self-confidence/stress, verbal, vocal, visual/non-verbal, audience management, context/preparation, presentation media/technology). This included a 7-point Likert scale for generalist/specialist and each topic, plus free text comments. Seven experts participated in two workshops in France (4 in Paris, 3 in Nantes), compensated at a fixed hourly rate. Topics most covered were self-confidence (5/7), speech preparation (4/7), verbal (3/7), vocal (3/7), visual (3/7), and audience management (3/7). Presentation media/technology was the least covered (2/7).

\subsection{Workshop procedure}

\begin{figure}
    \centering
    \includegraphics[width=.5\columnwidth]{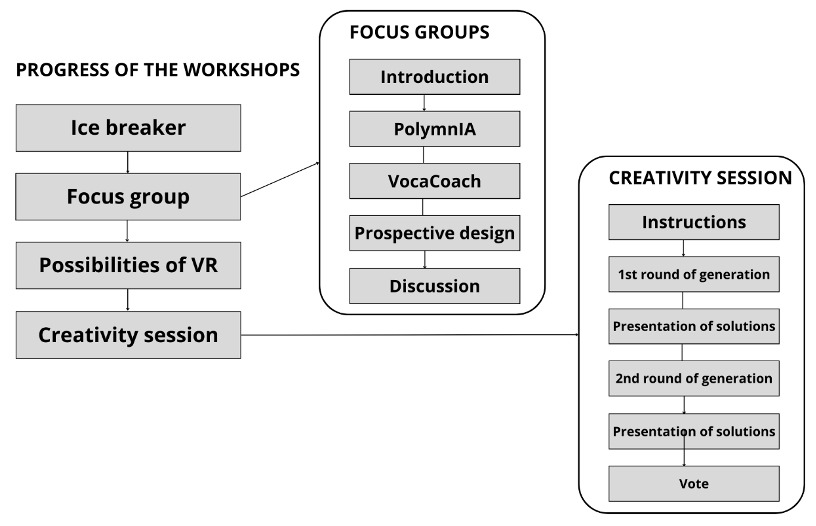}
    \caption{Overview of Study 2 workshops' procedure}
    \label{fig:design_workshop_procedure}
\end{figure}

This section details Study 2 workshop procedures, overviewed in Figure \ref{fig:design_workshop_procedure}.

\subsubsection{Ice-breaker}
We began workshops with an icebreaker: coaches chose animal cards (e.g., squirrel, bear, wolf, butterfly, whale) and imagined how they would perform public speaking imitating that animal. Beyond a simple icebreaker, this aimed to quickly engage trainers in discussing their public speaking visions.

% We started the workshops with an ice breaker step aimed at engaging the coaches in the workshop with a quick and enjoyable activity. For this ice breaker, we created a set of cards with photos of various animals (e.g. squirrel, bear, wolf, butterfly, whale ... ). At the start of the workshop, we invited the experts to choose an animal card presented face up on the table, and to imagine how they would perform public speaking if they wanted to imitate the animal on their card, for instance what it would be to speak like a butterfly, or a lion. Beyond the objective of a simple icebreaker, it was desired to quickly get to the heart of the matter, that is to say to get the trainers to talk about their visions of public speaking.

\subsubsection{Focus group introduction}
The focus group on AI-based public speaking training systems, along with the creativity session, formed the core of the workshop. We began by reminding experts of our objectives, then introduced a discussion framework based on classic brainstorming rules: no interruptions, positive listening, and no criticism. To guide the discussion, we presented three personas inspired by client types mentioned in the interviews:
% The focus group on AI-based public speaking training systems was the highlight of the workshop, along with the creativity session. We divided this into several stages. First of all, we reminded experts of our objectives. Then, we proposed a framework for managing discussions inspired by classic brainstorming guidelines: no interruptions, positive, judgment-free listening, refraining from criticism on others’ opinions. Finally, we presented three personas to facilitate common discussions, inspired by types of clients mentioned in the interviews:

\begin{itemize}
    \item Aude, a learner seeking self-confidence.
    \item Laurent: wants to learn conference speaking techniques.
    \item Agathe: a public speaking coach supporting Laurent and Aude.
\end{itemize}

\subsubsection{Presentation of PolymnIA and VocaCoach}\label{tools_description}
During the focus group, we presented commercial, AI-based public speaking training solutions to coaches to highlight market offerings. From representative AI tools (Section \ref{background_commercial}), we selected \textit{Polymnia} and \textit{VocaCoach} for our French-speaking panel due to their French interface (some coaches had limited English proficiency), accessibility, complementary design, user experience, and representativeness of modern AI-powered systems. By presenting actual usage scenarios, we aimed for expert opinions on pedagogical value, usability, and feedback nature. Table~\ref{tab:tool_analysis} provides a general description, but detailed interaction flows for Polymnia and VocaCoach are provided here due to their focus group importance.

% During the focus group, we then presented to the coaches a set of commercial, AI-based public speaking training solutions to highlight what is already been proposed in the market and what could be enabled. As presented in section \ref{background_commercial}, we had identified a set of representative AI-enabled tools. Considering our aim of organizing focus groups with our panel of French public speaking coaches, we selected only Polymnia and VocaCoach for inclusion in our study. Both systems are French-language, AI-powered, web-based public speaking training systems offering free and paid licenses,  aimed at the general public and requiring only a web browser and a webcam. The two systems were selected based on having French-language interfaces (some coaches in our panel claimed limited English proficiency) but also for their accessibility, their complementarity in terms of design approach, user experience, and their representativeness of the type of technological features, training and feedback paradigms present on modern AI-powered public speaking training systems that we aimed to assess. By confronting our expert participants with actual usage scenarios, we aimed to obtain experts' opinions on their pedagogical value, usability, and the nature of the feedback they provide.  While we provide a general description of these two systems in Table \ref{tab:tool_analysis}, given their importance in the focus groups, we provide here detailed descriptions of the interaction flow presented to users during a training instance with PolymnIA and VocaCoach.

\textbf{PolymnIA}\footnote[3]{\url{polymnia-france.com/}} is an online platform providing AI-enabled feedback on recorded speeches. Users record themselves via webcam/microphone, then receive a performance report. Trainees only specify speech duration; no themes or guidance are provided. During recording, the camera feed shows coloured overlays for detected behaviours (e.g., open postures). After recording, PolymnIA quickly offers a detailed report on verbal ("Fond"), vocal ("Forme"), and gestural ("Gestuel") aspects, with a general score out of 20 and various visualisations. A highly detailed 13-page document with extensive feedback categories can be downloaded. However, recorded video cannot be reviewed after analysis.

%Once the speech is finished, the system then quickly offers a detailed report on their speech covering verbal (``\textit{Fond}''), vocal (``\textit{Forme}'') and gestural (``\textit{Gestuel}'') aspects, all with a general score out of 20, and with a varied set of visualisations. Furthermore, the system provides an option of downloading a highly detailed 13-page document with extensive feedback categories. However, it is not possible to review the recorded video after it has been recorded and analysed.

\textbf{Vocacoach}\footnote[2]{\url{  https://vocacoach.fr/}}  is an online platform offering feedback based on user video recordings, with free testing and token purchases for further attempts. Similar to PolymnIA, users record speeches specifying estimated duration, but receive no preparation or recording advice. Unlike PolymnIA, Vocacoach only analyses vocal aspects.

Vocacoach’s feedback interface provides an overall vocal performance score (e.g., $80\%$) and a general feedback sentence (e.g., ``That’s a good speech! Certain points can still be improved...'' - note: author-translation). A spider chart shows five vocal dimensions: speech rate, tone, articulation, hesitations, and rhythm, each with a 1-5 performance estimate. The system summarises areas of high performance and areas needing work, plus speech duration. Users can click each vocal dimension for general information, description of system detection methods, and improvement tips, including pedagogical videos. For instance, the speech rate dimension offers examples of speakers with ``slow'', ``fast'', or ``right'' pace. 

%Furthermore, the user can click on each vocal dimension, which provides general information on this dimension. For example, for the hesitations dimension, Vocacoach provides a general definition, a description on how the system detects hesitations automatically, and a few tips on how to improve this specific aspect of speech; this includes, for some of the dimensions, pedagogical videos. For instance, for the speech rate the user is provided with examples of speakers who articulate their speech at the "right" speed, and others speaking too fast or too slow. 

\textbf{Demonstrations:} 
In workshops, we demonstrated PolymnIA by showing a pre-recorded two-minute speech and asking experts to note coaching remarks. We then shared PolymnIA’s feedback report and exhaustive document, allowing questions before demonstrating Vocacoach. A similar procedure was followed for Vocacoach: experts watched a pre-recorded speech, took their own notes on the speech, then we demonstrated Vocacoach and began the focus group. Unlike PolymnIA, Vocacoach allows reviewing the recorded performance.

After each demonstration, we discussed the presented system's design, encouraging participants to reflect on its feedback, interface, and potential use in their practice. System analysis is discussed in Section~\ref{results_study2}.

% In the workshops, we demonstrated the process of recording a video with PolymnIA to the experts, and showed them a pre-recorded video of a two-minute speech we uploaded. As we demonstrated the video, we asked the experts to write down any remarks they would have provided to the person training in the example video, i.e. as if they had been hired to coach them. After that, we shared with them the feedback report generated by PolymnIA on this video as well as the exhaustive document, and allowed them to ask a series of clarifying questions before moving on to demonstrate the second system, Vocacoach.
%  A similar procedure was followed to present Vocacoach: experts watched a two-minute speech recorded in advance and uploaded to Vocacoach, during which they were asked to write down any remarks as if they were about to coach the person in the recording. Then, we demonstrated Vocacoach and started the focus group discussion. Unlike PolymnIA, which does not record the video taken during the speech, Vocacoach offers to review the recorded performance. 

%  After each demonstration, we engaged in discussions over the design of the presented system. We encouraged participants to reflect on the feedback proposed by the systems, on their interface design, and how they might use such a system in their practice. The analysis of the systems is discussed in section \ref{results_study2}.

\subsubsection{Prospective Design} 
In addition to PolymnIA and Vocacoach, we presented several prospective AI-based public speaking training exercises. Authors contributed innovative ideas, including from academic research systems not yet commercialized. The goal was to invite participants to imagine future AI-based features beyond current commercial tools. We obtained 11 scenarios, selecting 5 most relevant and complementary to the presented systems.

% In addition to demonstrating our choice of two training systems, PolymnIA and Vocacoach, we presented to our participants several prospective designs of public speaking training exercices that could be implemented using AI-based technologies. To that end, each author  contributed ideas of innovative training exercises that could be enabled by technology, such as innovative research systems that had been proposed and studied in academia but not widely implemented in commercial systems. The goal was to invite participants to imagine other features, exercises or modules that could be implemented for AI-based public speaking training in the future, but which would have been absent today in our selection of commercial tools. We obtained a total of 11 potential prospective AI-based training system scenarios. Out of the 11, we chose the 5 scenarios that we felt were the most relevant to the workshops and the most complementary to the presented systems.

\noindent \textbf{Prospective Exercise 1: Management of silences}. An exercise to learn integrating silences by imitating famous speakers with lengthy pauses to practice and gain comfort.
% An exercise to learn how to better integrate silences into your speech could consist in imitating examples of speeches by famous speakers containing many lengthy pauses in order to practice and become more comfortable with using lengthier pauses. 

\noindent \textbf{Prospective Exercise 2: Posture and gestures}. Exercises to encourage more gestures and
open postures by performing speeches while imitating a silhouette, inspired by Schneider et al.~\cite{schneider2017presentation} and Barmaki et al.~\cite{barmaki2018embodiment}.
% To encourage users to make more gestures and adopt a more open posture, exercises could be imagined to perform a speech while adopting postures and gestures displayed by a silhouette, such as in the works of Schneider et al. \cite{schneider2017presentation} and Barmaki et al. \cite{barmaki2018embodiment}.

\noindent \textbf{Prospective Exercise 3: Insert emotions into speech}. Users purposefully add emotions to a prepared speech. Leveraging emotion recognition models (speech-based, multimodal), the interface provides feedback on recognized emotion type, checking if the prescribed emotion was successfully expressed.
% This would involve offering the user the objective of purposefully adding several emotions to a prepared speech. Leveraging emotion recognition models (e.g. speech-based, or multimodal), the interface would provide feedback on the type of emotion recognized and check if the participant successfully expressed the prescribed emotion.

\noindent \textbf{Prospective Exercise 4: Peer feedback}. A platform for a community of users to provide peer feedback, similar to RocSpeak \cite{Zhao2017}.

\noindent \textbf{Prospective Exercise 5: Virtual audience interaction}.
Managing audience interaction was highlighted as an important skill by public speaking specialists. This exercise, based on Virtual Reality (VR), displays a simulated audience in a virtual room. Prior research extensively investigated virtual audiences \cite{Batrinca2013,chollet2015exploring,van2020impact,chollet2022training,glemarec2023towards,ristorcelli2024impact}. In line with our focus on AI-based feedback, we presented concepts for visualising user behaviour in such contexts—e.g., post-training reports or gaze heatmaps. This was illustrated through a live demonstration of the VR system developed by Ristorcelli et al. \cite{ristorcelli2024impact}.

% Being able to manage interactions with an audience was considered important for public speaking specialists. We included one prospective exercise to train these skills. This exercise is designed within the framework of a Virtual Reality interaction, in which a virtual room is displayed with virtual characters representing the audience. Much prior research has investigated the concept of virtual audiences \cite{Batrinca2013,chollet2015exploring,van2020impact,chollet2022training,glemarec2023towards,ristorcelli2024impact}. To fit with the broader theme of our research on the concept of AI-based public speaking feedback, we presented our participants with concepts where users' behaviour could be visualised or analysed in the context of a virtual audience interaction, such as through post-training reports or visualizations, such as gaze heatmaps over the virtual audience. We illustrated the capabilities of Virtual Reality and  virtual audiences to simulate public speaking through a live demonstration of the system developed by Ristorcelli et al. \cite{ristorcelli2024impact}.

\subsubsection{Creativity Sessions}
To conclude the design workshops, we conducted creativity sessions using the nominal group technique \cite{delbecq1971group,dunham1998nominal,harvey2012nominal}. Participants worked individually, then pooled thoughts. We presented the process and the central question: ``In the future, what system(s) would be truly efficient for learning public speaking?'' Secondary considerations included: ``For whom? How often? With whom? What key moments/activities?''

% To conclude the design workshops, we organised creativity sessions. For this purpose, we opted for the nominal group technique method \cite{delbecq1971group,dunham1998nominal,harvey2012nominal}. The nominal group technique consists of having each participant work individually, then pooling their thoughts. When starting this step of the workshops, we first presented the creativity session process to the participants, then expressed this central question: ``In the future, what system(s) would be truly efficient for learning public speaking?''. After this main question, we encouraged them to think of other secondary considerations: 

Materials (A3 sheets, post-its, pens, stickers) were prepared. For the individual step, participants had five minutes to write as many uncensored ideas as possible. After, each presented their proposals. In the second workshop, an additional round allowed experts twenty minutes to produce a final individual proposal, incorporating peers' ideas, which they then presented. Finally, participants voted on their preferred solution and discussed it.

% Materials were prepared in advance for organizing this session: these questions were posted on A3 sheets attached to the room walls. Participants had post-it notes, A3/A4 sheets, pens, and stickers at their disposal. 
% For the individual step, the participants were given five minutes to write their answers to the above questions; they were encouraged to write as many ideas as possible with as little self-censorship as possible. Before starting the session, we made sure participants understood the objectives and allowed them to ask any questions. After the five minutes, each participant presented their proposals to the group. As more time was available, we were able to add an additional round to the second workshop allowing the experts were asked to produce a final proposal, this time drawing also on the proposals presented previously by their peers. They had 20 minutes to produce this final individual solution, which they presented then to the group. Finally, participants voted on their preferred solution and discussed it in a final conversation.

\subsection{Data Analysis and Key Insights}
\label{results_study2}
Similarly to Study 1, we used inductive thematic analysis to analyse the focus group transcripts. We highlight key insights from coaches during workshops, covering their analyses of system
design, prospective exercises, and suggestions from creativity sessions.

\subsubsection{AI-based training systems afford useful, novel self-guided training opportunities}

\textbf{Supporting repeated public speaking rehearsals is a key benefit.}   Trainers were very receptive to using systems like Vocacoach or PolymnIA, especially for independent practice between sessions and providing basic feedback on core skills. Coaches overwhelmingly agreed such systems are ideal for enabling regular practice, allowing learners to practice a given speech multiple times.
% The trainers were very receptive to the idea of using systems such as Vocacoach or PolymnIA in their practice, in particular to enable learners to practice independently between sessions, but also to provide them with regular, basic feedback on core skills.
% Coaches overwhelmingly agreed that it is very important for trainees to regularly practice, and that such systems provide an ideal platform to enable this regular practice. Our participants repeatedly mentioned being very favorable to the fact that learners could use such training applications to practice a given speech multiple times
\begin{quote} \textit{``I am amazed to see that speakers often learn their lectures by heart. You have to rehearse speeches, and you have to practice hearing yourself. And with software, it's great because [trainees] understand that they have to rehearse [a speech] out loud.''} (W1-P4) \end{quote}

\textbf{The neutrality of an automated system can defuse emotional barriers to training.} Training in the safety of one’s home with a neutral computer solution can disinhibit trainees: \begin{quote} \textit{``Users can train as much as they want. […] what people appreciate [with AI-based feedback] is being judged by a machine and not by human beings. There is a gamified aspect: you can track your progress. This type of software allows learners to practice at home. Sometimes that's what's missing for progression.''} (W1-P2) \end{quote} 

W1‐P4 agrees, noting fewer emotional "risks" with a machine. Another advantage is objective speech comparison over time. W1‐P4 recounts improving their rhythm by self-recording their speech:
% W1-P4 agrees, expressing the idea that with a machine, there are fewer emotional "risks" of a trainee losing motivation due to them fearing feedback or interaction with peers or coaches. Another advantage of using such a platform is in comparing speeches objectively, over time. W1-P4 talks about her experience improving following a remark about a lecture she gave repeatedly at conferences: she always had the same rhythm when giving this lecture, and she thought she had to break this habit to regain authenticity as a speaker. She practiced a lot to break this rhythm:

\begin{quote} \textit{``I recorded myself, and listened to myself again, and again... There, the machine tells me: at this moment, you accelerated... I [can choose to focus or not on] this analysis. But I still find it interesting.''} (W1-P4) \end{quote}

\textbf{Training systems can take on time-consuming exercises, allowing coaches to focus on higher-level concerns.} W1‐P4 is enthusiastic about AI tools enabling independent practice of speaking techniques, delegating tasks they might not cover or would assign to a system. \begin{quote} \textit{``The tool represents the technical grid that is necessary […]. The software is interesting, by its neutrality, its technical, emotionless side. We [humans] are too connected to each other and do not want to offend [our trainees], to not break their motivation. [An AI-based feedback tool] is external, technical, but it gives indications, and as a coach it can save me time, time to work on the sensitivity of my client. These tools […] can free me from the basics.''} (W1-P4) \end{quote} 

\noindent\begin{center}
    \fbox{%
    \parbox{.9\columnwidth}{%
        Insight 1: Experts find value in AI-powered feedback systems' capacity to enable repeated, frequent rehearsals in a low-stakes, neutral environment, minimising trainees' emotional risks. This can free up coach time to focus on higher-level pedagogical, personalised concerns. 
    }%
}
\end{center}

From this perspective, an AI-powered analysis tool like PolymnIA, providing neutral, technical feedback, could transform the trainee-coach dynamic. It frees the trainer to focus on higher-level aspects: transmitting emotions, uncovering authentic speaking styles, and identifying activities to improve these areas.

\subsubsection{The complexity of designing public speaking feedback}

A key focus group topic was AI-feedback system design. All coaches agreed it is difficult to gauge the right feedback level for trainees, and both presented systems often missed the mark. We identified several subthemes from these discussions.

\paragraph{Public speaking analysis criteria should be carefully selected and organised}

\textbf{Feedback elements need to be understandable and carefully selected according to the speech context and user goals.} 
Feedback elements need to be understandable and carefully selected according to speech context and user goals. Most coaches believed PolymnIA reports were too detailed, inducing cognitive load leading to users not knowing what to focus on. According to W2‐P7, it is difficult for learners to parse through numerous remarks, applying all of them:
\begin{quote} \textit{``I give 2-3 comments maximum on trainees’ performances, otherwise they will not remember anything and it will be useless''.}  (W2-P7)\end{quote} Furthermore, experts weren't sure users would correctly understand all feedback items, as some covered complex or obscure public speaking notions without providing adequate explanations or improvement suggestions: \begin{quote} \textit{``For feedback to be relevant, systems should present short videos to explain key concepts. There, [feedback] remains abstract and we [can] do nothing with it.''}  (W2-P6)\end{quote} 

According to trainers, some system criteria were problematic, being either irrelevant depending on the situation or poorly structured in terms of importance or key categories. First, some feedback categories may not be relevant depending on speech type, message, context, audience, and speaker intentions. W1‐P1 gives the example of Polymnia criteria "Use of Gerund Terms": \begin{quote} \textit{``In any speech, intended for any audience, […] we cannot affirm that in all situations we need the same [evaluation] criteria.''} (W1-P1) \end{quote} W1-P4 adds that a public speaking performance cannot be analysed in isolation, but only in relation to its goals and the specific situation it is delivered in: \begin{quote} \textit{ ``Does the system take into account the trio (1) My intention, my objective, (2) The occasion, the context (3) The message?''} (W1-P4) \end{quote} W1‐P4 concedes that trainees might select themselves, from a larger list of AI-analysable criteria, which ones to consider and disregard. This may be particularly relevant in tandem with a coach, who could identify particular focus criteria for a trainee. Second, according to W1‐P1 and W1‐P3, feedback structuring is confusing with various criteria presented side by side, but not in order of importance or by theme: \begin{quote} \textit{``There are several types of criteria mixed together''} (W1-P1) \end{quote} They highlight PolymnIA report's successive five items: (1) Most frequently used words, (2) Speech theme orientation, (3) Marketing orientation, (4) Gesture analysis chart (5) Intonation analysis. To them, some remarks concern high‐level speech elements, while others concern very specific, low‐level behavioural elements.

Compared to PolymnIA, participants thought Vocacoach seemed more suited to independent user training due to more concise, high‐level feedback with fewer, more understandable and actionable elements. They fear PolymnIA would seem overwhelming with its exhaustive, technical feedback. Participants agreed that an ideal application should be configurable by public speaking coaches to present only feedback relevant to learners at particular training journey moments: \begin{quote} \textit{``As a coach who supports a trainee [in their learning journey], I should be able to choose the feedback that appears in the report.''} (W1-P2) \end{quote}

\textbf{Feedback valence should be generally neutral; encouraging to beginners.} Interestingly, the two systems were seen as opposed in positioning—PolymnIA being too detailed, Vocacoach too surface‐level. A participant noted Vocacoach's feedback was generally positively biased, potentially problematic if learners used the system before in‐person coaching sessions. For instance, trainees may receive highly flattering feedback, unduly improving their public speaking confidence. W2‐P7 notes: \begin{quote} \textit{``What scares me would be someone who uses [such a system] before we see each other, and it calls into question my advice. For the ego, it’s good but [it can be counter-productive for speaking skills]''.}  (W2-P7\end{quote} Still, this design choice of simplicity and positivity in providing feedback could be of interest to some users. W1-P3 notes that simple, generic recommendations can be good starting points: \begin{quote} \textit{``What I find interesting [in Vocacoach] is the more simplified aspect. In particular, the fact that they explain a little about what is generally expected, for example, on rhythm. (…) Sometimes you have to repeat the obvious, even to experts.''}  (W1-P3)\end{quote} TThus, systems like Vocacoach would be primarily useful for absolute beginners starting their public speaking improvement journey: \begin{quote} \textit{``It depends on the [trainee's] level. The feedback [of Vocacoach], for me, is intended for someone who is just starting out.''}  (W1-P1)\end{quote}

\noindent\begin{center}
    \fbox{%
    \parbox{.9\columnwidth}{%
        Insight 2: Experts highlight the complexity in designing public speaking feedback. It should be carefully selected, limited to a few key elements, organised logically and hierarchically, and tailored to the trainee's goals and the speech context; it should be understandable and actionable; it should generally be neutral, although beginners may benefit from encouragement.
    }%
}
\end{center}

\paragraph{Feedback should address higher-level communicative goals}

\textbf{Lower-level public speaking behaviour should be analysed insofar as it conveys higher-level meaning.} Participants observed that many PolymnIA and Vocacoach feedback items focus on lower‐level behaviour trends while missing high‐order concepts of good public speaking performances. Commenting on pause and silence management feedback, W1‐P4 commented: \begin{quote} \textit{``We choose to be silent because there is meaning in silence, and [AI-feedback] does not say this.''} (W1-P4)\end{quote}  W1-P4 provides another example of missed high-level criteria: \begin{quote} \textit{``Speaking speed doesn't mean much in itself. Perhaps someone who speaks very quickly would have an interest in taking longer silences. [The interface] is missing the criteria of ‘energy’: when you speak, you need energy.'' } (W1-P4). \end{quote}

One coach expressed that complex evaluation criteria, such as enthusiasm, commitment or energy, are missing: \begin{quote} \textit{``Here in the example feedback you show us […], I'm missing something of the enthusiasm, of the commitment [of the speaker]'' } (W1-P4). Participant W1-P2 explains how higher-level criteria such as ``commitment'' are eventually expressed through the speaker’s non-verbal behaviour: \textit{``it’s in the look, it’s in the gestures.''} (W1-P2) \end{quote} 

Are low‐level behavioural characteristics insufficient since what truly matters is higher‐level criteria? One participant mentions that the elements are intrinsically linked and what matters is providing clear relationships between behavioural indicators and higher‐level criteria: \begin{quote} \textit{``For [evaluating a speaker’s] energy, [an ideal system would] look at the variation in expressions of the user. Does the user change their state a lot.''} (W1-P2) \end{quote} 

\textbf{The focus on lower-level public speaking behaviour feedback risks trainees conforming to  inauthentic styles.}  Several coaches questioned whether focusing on specific, objective behavioural criteria could encourage conformity. They argued that authenticity is essential and conforming to stereotypical speaking can be harmful: \begin{quote} \textit{"Those who train [with such a simplistic view] can acquire a sort of robotic style, which will ultimately deprive them of the nuance that coaching can provide."} (W1-P4); \textit{``The problem with using a grading grid is that we break the naturalness by focusing on precise criteria. In addition, speakers will all become the same.''}  (W2-P6); \textit{``Be careful not to make speakers become too 'smooth', because that is not authentic.''} (W2-P6).\end{quote}  

W1-P3 evokes risks that exist in focusing on behaviour and technique. They give examples of speakers who can be very good technically, but who do not capture and engage the audience well. W1-P4 adds: \begin{quote} \textit{``When you’re not great and all you have is technique, it’s boring.''} (W1-P4) \end{quote}

\noindent\begin{center}
    \fbox{%
    \parbox{.9\columnwidth}{%
        Insight 3: Experts highlight how feedback should target higher-level dimensions such as speakers' conveyed energy, enthusiasm, commitment. Lower-level behavioural feedback may not be able to be acted upon, and runs the risk of encouraging trainees to conform to a standard, inauthentic style.
    }%
}
\end{center}

\subsubsection{Public speaking training activities need contextualization and clear instructions}

\textbf{Public speaking training activities should be based on clear goals.} Coaches evoked an issue regarding both systems related to the context in which they would be used. In the demonstrated versions, users were not given or asked any specific speaking goals or contexts: the only information required was the intended speech duration. Beginners or unprepared users may not have prepared topics or may need more guidance to start training. Further, coaches pointed out that in the absence of key motivation behind a speech, performance can appear flat as there is no key message to convey or communicative intention driving the trainee. Thus, it is counterproductive to improving speaking skills without specific context or topic: \begin{quote} \textit{``In this system, the person is completely left to their own devices. During coaching sessions, we provide [basic contextual elements] to at least start the exercise. Indeed, without a situation, they can completely miss the exercise [i.e. exhibit poor speaking performance]. And if the results are too bad, it risks confirming [their belief] that they are bad at public speaking.''} (W2-P7) \end{quote}

W1‐P2 highlights that setting training goals is a key preliminary step before engaging in any training aspect, which helps contextualise and give meaning to activities, as well as motivate learners towards clear objectives:\begin{quote} \textit{ ``Every time we gave a speech [to a public speaking training group she used to lead], we set the objectives of the speech. And we have someone who evaluates  [the speaker] against these specific objectives''} (W1-P2) \end{quote} 

A specific section of the PolymnIA report provides an automated analysis of the speech theme. In one generated report, the system provides an analysis of one example speech through an analysis of the speech transcript, e.g.:``\textit{Your speech’s orientation would lean towards the Literary and Philosophical domain. The below list of terms in your speech were the most relevant in this area: Imagine, Book, Who, Time, Unique, Being}''. This element was noticed by coaches as a particularly interesting piece of feedback: \begin{quote} \textit{``The AI [should] be able to detect [whether this was] the user's intention, otherwise it is missing the point.''} (W1-P1) \end{quote} According to this participant, the system's ability to analyse speech theme is interesting but only relevant if this analysis relates to the trainee's goal. For instance, users could verify whether they managed to convey the right theme and adjust their speech accordingly if a discordant theme appears predominant. This reinforces that context and goal in which trainees engage with particular activities matters and should inform provided feedback.

\textbf{Clear instructions should be provided to encourage trainees to meaningfully engage with training. }
Another important discussion item concerned the physical context in which public speaking was realised. Both PolymnIA and Vocacoach did not provide information or guidance on how to record oneself. W2‐P7 was surprised that users were not asked to stand up or given instructions on camera distance:  \begin{quote} \textit{``Sitting cancels out part of the exercise''} (W2-P7) \end{quote}
Similarly, this lack of contextualization and instructions was seen by W1‐P3 as leading to key mistakes, such as reading speeches from written notes, as observed in examples shown: \begin{quote} \textit{``I noted an essential thing [in this example] that I do not find in the feedback: from the intonation, the rhythm, I have the impression that [the speaker] was reading something. And so there was a rhythm due to the reading which […] lacked naturalness. These are essential things that I address all the time [in their practice]''} (W1-P3) \end{quote} The absence of key physical and contextual instructions on performing training activities can lead to counterproductive practice. A similar element pointed out was the interface design of the training activity itself. When recording practice speeches, trainees face their screen and can see live webcam capture; on PolymnIA, during speech, the performance video is superimposed with coloured squares highlighting detected gestures and facial expressions. W2‐P7 notes: \begin{quote} \textit{``Seeing yourself, you study yourself and you are less natural.''} (W2-P7).  W2-P5 agrees:\textit{``[Trainees] risk learning to speak in front of a computer [instead of in front of an audience]"} (W2-P5) \end{quote}

An obvious design issue noted by W2-P5 regarding PolymnIA was that the detailed feedback was unverifiable and difficult to contextualise by users because they could not review their video, only their transcript. Feedback systems should provide mechanisms to review past presentations in order to reflect on feedback focused on behaviour, ideally time-aligned to easily retrieve segments of poor or good performance. \begin{quote} \textit{``The document gives advice but we no longer know what we did [at that time]. It remains very abstract''.}  (W2-P5)\end{quote} 

\textbf{Systems should contain activities for knowledge-acquisition and exercises for practicing essential skills.} One participant mentions that applications should provide guidance and knowledge acquisition activities before actual training rehearsals followed by public speaking analysis. \begin{quote} \textit{``They [should] provide a training space. I think that for people who train all alone, it's a very good start.''} (W1-P2) \end{quote} They argue that undertaking exercises to perfect certain behavioural techniques and performing full speech practice are very different activities and should be separated with different activities: e.g. video examples and training exercises with specific feedback for practicing techniques, separated from global speech rehearsal modules with holistic feedback as presented by PolymnIA and Vocacoach.

\noindent\begin{center}
    \fbox{%
    \parbox{.9\columnwidth}{%
        Insight 4: Experts described how seemingly obvious instructions and contextual elements need to be provided before public speaking practice: users should specify clear goals and objectives helping define the training activities, or be provided with clear contexts including speech topics and audience descriptions; trainees should be given physical instructions such as standing and performing their speech as in real conditions; finally, systems should support knowledge acquisition and specific behavioural exercises as separate activities from public speaking rehearsal followed by holistic feedback.
    }%
}
\end{center}

\paragraph{Training activities should adapt to the speaker's personal speaking style}

\textbf{Public speaking feedback should foster the expression of trainees' authentic styles.} Earlier, we described how participants highlighted that low‐level feedback risks leading trainees to adopt conformist public speaking styles. Related to this, participants expressed that autonomous learning systems such as PolymnIA and Vocacoach should help speakers identify their personal public speaking style and tailor activities to strengthen it. Participant W2‐P7 mentioned: \begin{quote} \textit{``I don’t like [speakers] who are too formatted, who make certain gestures because they have been told to do it that way. This irks me because I no longer see a person, I only see technique. In public speaking, what we want is the [authentic] person.''} (W2-P7) \end{quote} Thus, the training should adapt to the style of the speaker, building on it instead of forcing trainees to adopt a specific style: \begin{quote} \textit{"I don’t ask [of my clients] that they be like me. When I talk about energy in my coaching, I am going to take their own energy into account."} (W1-P4) \end{quote} 

A crucial design question is how to develop public speaking skills with AI‐based interfaces, necessarily trained on certain reference public speaking training data, while allowing trainees to maintain authentic styles. Here, we see disagreement on achieving this through training course organization. W2‐P5 believes it's fine to start with learning formatted, rigid techniques before emancipating oneself: \begin{quote}
\textit{``We must first learn how to speak in an ``academic'' way, and then [learn to] display our own style.'' } (W2-P5)
\end{quote} Disagreeing, W2‐P6 argues you must first learn to be natural and authentic before perfecting your style through specific techniques. W1‐P1 frames their coaching similarly: instead of focusing on correcting mistakes, they focus on learner strengths, taking authentic speaking style as starting point, helping improve them further: \begin{quote} \textit{``As a trainer, […] I rely on my client’s rhythm, style, tone. […] So if I impose you [a rigid set of speaking style criteria], even if I think it's correct, I'm not taking your nature into consideration.}  (W1-P1)\end{quote}

Another coach indicates that regardless of overall organization, training activities should be organized depending on trainee level and that exercise complexity, feedback, and overall training activities should augment progressively: \begin{quote}
\textit{``The more you increase in level, the more you must increase the finesse of the feedback''} (W1-P1)
\end{quote}

\noindent\begin{center}
    \fbox{%
    \parbox{.9\columnwidth}{%
        Insight 5: Public speaking training systems should be personalised to trainees according to two key elements: their current level of experience in public speaking, and their authentic style of speaking and behaviour.
    }%
}
\end{center}

\section{Discussion}\label{discussion}

%In the next section, we summarize the main findings of our study, propose guidelines and perspectives for designing future AI-based public speaking training systems and discuss the limitations of our work.

% % % % % Main Topics of the discussion to by summarized : 
% {Summary of coaches’ insights on Systems of Automated Public Speaking}
% {Automatic Public Speaking Assessment Systems From the Literature Perspective.}
% \subsubsection{Lack of Personalisation and Context Sensitivity}
% \subsubsection{Neglect of Emotional and Psychological Dimensions}
% \subsubsection{Surface-Level Performance Metrics vs. High-Level Communicative Goals}
% \subsubsection{Insufficient Support for Structured Learning and Reflection}
% \subsubsection{Lack of Cultural and Linguistic Sensitivity}
% {Limitations of our studies}

%\subsection{Summary of coaches’ insights on Systems of Automated Public Speaking}

%Developing public speaking skills requires regular practice. With this key consideration in mind, a
A first key insight from our focus groups is that being able to practice easily, on-demand, using a judgment-free automated system was found to be a very useful feature for all our participants. Two types of training modalities were identified : 1) holistic speech practice and 2) purposeful pedagogical training activities (such as managing silences, articulation, voice modulation and breathing exercises). Our sample of experts highlighted the valuable use of an AI-feedback system as a complement tool to focus on performing specific behavioural exercises.%, thus leaving to the coaches higher-level feedback on the communication cues ( personal style, energy, emotions, authenticity ...). 

The analysis of current automated public speaking training systems, combined with expert interviews, reveals a notable disconnect between existing technological implementations and the pedagogical foundations articulated in the public speaking literature. To ensure the reliability of AI- based feedback for automated training activities, the experts accentuated the need to consider the user's goal, profile and situational context to contextualize the speech practice (personal/ professional context, the audience size, ...). The coaches also highlight the need to provide clear instructions regarding physical and contextual aspects, and to produce organized, selected, and actionable feedback. Finally they also stressed the need to tailor the feedback to the speaker's idiosyncratic style in order to avoid the fitting into a standardized manner of speaking. This design contrasts sharply with the recommendations found in authoritative public speaking manuals such as Lucas's \textit{The Art of Public Speaking} \cite{lucas2020art}, Duarte's \textit{Resonate} \cite{duarte2010resonate}, and Toogood’s \textit{The New Articulate Executive} \cite{toogood2010articulate}. These works emphasize that effective public speaking is inherently audience-centered, culturally situated, and purpose-driven.

A striking omission in all analysed systems is also the absence of support for public speaking anxiety, despite its prevalence among learners. This is particularly concerning, as all interviewed coaches identified anxiety management and cognitive restructuring as the foundational steps in their training. They noted that confidence must precede technique, and many trainees require support in overcoming feelings of illegitimacy and low self-worth before engaging meaningfully with performance skills. This pedagogical stance has been also corroborated and discussed in previous research works \cite{lucas2020art,toogood2010articulate}.

%This pedagogical stance is corroborated by Lucas \cite{lucas2020art}, who dedicates extensive discussion to strategies for transforming stage fright into ``positive nervousness,'' offering evidence-based techniques such as visualization, deep breathing, and reframing negative thoughts. \cite{toogood2010articulate} also proposes steps for fear overcomming and point out the difference between the fear that wakes up the mind and the anxiety and stage fright. They also propose to move towards positive thinking and reverse the internal view towards the fun and humanistic aspect of the performance. The lack of these components in digital training environments constitutes a critical oversight, undermining their potential to scaffold early-stage learners.

%\subsubsection{Surface-Level Performance Metrics vs. High-Level Communicative Goals}

Our review also reveals that most systems prioritise numerically quantifiable aspects of delivery, such as word count, average pitch level, pausing count, or hand gesture detection. While essential, these features can not fully represent such aspects as speaker authenticity, emotional resonance, or the capacity to inspire. Coaches expressed concern that this overemphasis on technical correctness could lead to robotic and homogenised speech patterns, disconnected from real audience impact.
This critique echoes the frameworks presented in \cite{duarte2010resonate}, where the emotional and structural depth of a performance is central to its effectiveness. %Duarte’s narrative-driven model encourages speakers to craft transformational experiences for their audiences, not just to perform flawlessly. 
Similarly, Lucas \cite{lucas2020art} and Toogood \cite{toogood2010articulate} both highlight that rhetorical effectiveness arises from a synthesis of content, form, and emotional delivery—dimensions difficult to capture through surface metrics alone. 

%\subsubsection{Insufficient Support for Structured Learning and Reflection}

%While some systems offer isolated drills or basic educational content, they 
Moreover, current systems lack integrated curricular pathways or coach-configurable progression plans. Coaches emphasized the importance of guiding learners through preparatory stages, deliberate practice, and iterative feedback loops. Furthermore, they noted the value of video review paired with reflective prompts to consolidate learning. These instructional needs are directly aligned with the didactic structures proposed in the literature \cite{lucas2020art,toogood2010articulate} .
%Lucas \cite{lucas2020art} outlines a scaffolded model of speech preparation, including topic selection, audience analysis, speech structuring, and visual aid integration—components supported by reflective exercises and guided practice. Toogood \cite{toogood2010articulate} similarly promotes the \textsc{POWER} framework (Punch, One theme, Windows, Ear, Retention) to organize message delivery while enhancing memorability and emotional engagement. They further advocate for experiential learning exercises (e.g., the ``8-second drill'') that simultaneously build fluency and reduce anxiety through structured exposure. 
The gap between these comprehensive educational models and the fragmented structure of current systems highlights a missed opportunity for meaningful learning support.

%\subsubsection{Lack of Cultural and Linguistic Sensitivity}

Lastly, current systems %offer limited functionality for non-English users and 
do not account for cultural differences in rhetorical norms, non-verbal communication, or audience expectations. This shortcoming was noted both in our own system review and by several coaches working in multilingual or multicultural contexts. This concern is explicitly addressed in the literature. Lucas \cite{lucas2020art} devotes his attention to intercultural communication, pointing out that gesture interpretation, vocal delivery norms, and even rhetorical preferences vary significantly across different cultures. There is no universally accepted standard for effective speech delivery \cite{lucas2020art}. For instance, in terms of speaking rate (e.g.,  120 to 150 words per minute typically for US), historical examples illustrate this variability: Franklin Roosevelt spoke at a deliberate 110 words per minute, John F. Kennedy at a rapid 180, and Martin Luther King Jr. ranged from 92 words per minute at the start of his ``I Have a Dream'' speech to 145 by its conclusion.

\section{Limitations}
Although our two studies provide valuable insights on public speaking experts' practice, trainees, and views on public speaking training (Study 1), as well as on their perspectives of commercial public speaking training tools (Study 2), our research remains limited in scope regarding the diversity of reviewed systems and explored dimensions. Our in-depth analysis of two commercial public speaking training tools (PolymnIA and VocaCoach), while representative of distinct approaches, does not encompass the full spectrum of emerging system functionalities - particularly those related to linguistic, cultural, or contextual variability in oral communication. For instance, many of the listed systems have since deployed Large Language Models in their tools, offering new capabilities such as analyses and recommendations on  speech transcripts. Moreover, while the number of expert coaches involved in our studies allowed for a rich qualitative analysis and perspective, future investigations with larger and diverse panels could further enrich the understanding of experts' expectations, such as studies evaluating the actual use of these tools by experts in their practice. These considerations underline the relevance of pursuing broader and longitudinal studies to capture additional dimensions of public speaking training systems' potential and validate evolving feedback design principles in a wider range of usage contexts. 

\section{Conclusion}\label{conclusion}

This paper reports on two participatory design studies with public speaking coaches. Study 1 involved interviews on expert practice, trainees, and views on public speaking training. Study 2 consisted of two design workshops evaluating experts’ opinions on AI-based public speaking training systems.

Analysis of Study 1 interviews identified key aspects of public speaking coaching: 1) trainees’ needs (often negative self-evaluations, needing confidence support), 2) public speaking performance quality dimensions (content, form, emotions), and 3) coaching pillars (preparation, stress management, delivery).

Study 2 revealed experts’ perspectives on AI-based feedback systems. Overall, experts view AI-driven tools as useful, especially for rehearsal and behaviour tracking. However, they emphasize expert supervision in selecting/prioritizing feedback, adjusting it to user profiles and goals, and preserving individual authentic speaking styles.

A main finding of our workshops is that a key complexity for the development of automated public speaking tools lies in the design of feedback. Striking the right balance in the level of detail of feedback is crucial; while detailed analysis of behaviour may support expert speakers with highly specific goals, it can overwhelm beginners. The tone of feedback is equally important; while automated systems tend to focus on the weaknesses of the trainee, expert coaches often build on strengths to support learning and confidence. An effective tool should integrate both positive reinforcement and highlight areas of improvement. Moreover, there is a need to bridge low-level behavioural cues with high-level communication criteria. Isolated metrics such as gesture frequency or pause length offer detailed feedback but with limited value unless it is linked to broader high-level communication qualities such as energy, naturalness, enthusiasm and emotional engagement. The relationship between these layers is essential for actionable feedback. 

Another complexity stems from the contextualization of the feedback and training activities. Generic AI-based recommendations in the systems we presented to our participants often failed to address the specific goals, audience type, or  format of a public speech. Besides, certain behaviours or stylistic choices must be interpreted as relative to the speaker's level of expertise, the context of the training. Further, experts cautioned that over-standardized feedback risks eroding speakers' authentic speaking styles by promoting uniform behaviour.

Based on these findings, we offer several design principles for future public speaking training systems: personalise feedback according to the speaker style, profile and training stage, tailoring analysis to the context of the speech delivery, providing clear instructions for contextualising training activities, and last but not least developing models able to establish the relationship between behavioural patterns and higher-level performance criteria over time. Finally, the integration of large language models offers promising perspectives for future automated training systems, particularly for contextual understanding and simulated interactions. However, their use should remain aligned with evidence-based pedagogical goals and personalised support.

\section*{Author contributions}

\textbf{Nesrine Fourati:} Investigation, Formal analysis, Writing - Original Draft, Writing - Review \& Editing, Funding acquisition. \textbf{Alisa Barkar:} Formal analysis, Writing - Original Draft, Writing - Review \& Editing. \textbf{Marion Dragée: } Writing - Original  Draft, Writing - Review \& Editing. \textbf{Liv Danthon-Lefebvre:} Conceptualization, Methodology, Investigation, Formal Analysis, Data Curation, Project administration. \textbf{Mathieu Chollet:} Conceptualization, Methodology, Investigation, Formal analysis, Writing - Original Draft. Writing - Review \& Editing, Supervision, Project administration, Funding acquisition

\section*{Acknowledgments}
We are immensely grateful to the public speaking coaches and experts that participated in our studies. This research was partially funded by the French Agence National de la Recherche (ANR) under the REVITALISE grant (ANR-21-CE33-0016-02).

\section*{Conflict of interest}

The authors declare no potential conflict of interests.

\bibliographystyle{unsrt}  
\bibliography{references}

\end{document}